\documentclass[prd,tightenlines,nofootinbib,superscriptaddress,showkeys]{revtex4}

\usepackage{amsfonts,amsmath,amssymb,amsthm,bbm,hyperref, mathrsfs }
\usepackage{color,psfrag}
\usepackage[dvips]{graphicx}
\usepackage{slashed}

\newcommand{\und}{{\text{ and }}}
\newcommand{\csch}{{\text{csch}}}

\newcommand{\Ei}{{\text{Ei}}}

\newcommand{\e}{{\text{e}}}
\renewcommand{\i}{{\text{i}}}

\begin{document}

\title{Fractional Effective Action at strong electromagnetic fields}

\author{{\bf Hagen Kleinert}\email{h.k@fu-berlin.de}}
\affiliation{ICRANet, Piazzale della Repubblica 10, 65122 Pescara, Italy}
\affiliation{Institut f\"ur Theoretische Physik, Freie Universit\"at Berlin, Arnimallee 14, 14195 Berlin, Germany}
\author{{\bf Eckhard Strobel}\email{eckhard.strobel@gravity.fau.de}}
\affiliation{ICRANet, Piazzale della Repubblica 10, 65122 Pescara, Italy}
\affiliation{Dipartimento di Fisica, Universit\`a di Roma "La Sapienza", Piazzale Aldo Moro 5, 00185 Rome, Italy}
\affiliation{Universit\'e de Nice Sophia Antipolis, 28 Avenue de Valrose, 06103 Nice Cedex 2, France}
\author{{\bf She-Sheng Xue}\email{xue@icra.it}}
\affiliation{ICRANet, Piazzale della Repubblica 10, 65122 Pescara, Italy}
\affiliation{Dipartimento di Fisica, Universit\`a di Roma "La Sapienza", Piazzale Aldo Moro 5, 00185 Rome, Italy}

\date{\today}

\begin{abstract}
In 1936, Weisskopf showed \cite{Weisskopf1936} that for vanishing electric or magnetic fields the strong-field behavior of the one loop Euler-Heisenberg effective Lagrangian of quantum electro dynamics (QED) is logarithmic. Here we generalize this result for different limits of the Lorentz invariants \(\vec{E}^2-\vec{B}^2\) and \(\vec{B}\cdot\vec{E}\). The logarithmic dependence can be interpreted as a lowest-order manifestation of an anomalous power behavior of  the effective Lagrangian of QED, with critical exponents \(\delta=e^2/(12\pi)\) for spinor QED, and \(\delta_S=\delta/4\) for scalar QED.
\end{abstract}


\maketitle


\section*{Introduction}
In 1931 Sauter \cite{Sauter1931} and four years later Heisenberg and Euler \cite{Heisenberg1936} provided a first description of
the vacuum properties of QED. They identified a characteristic scale of strong field
$E_c=m_e^2c^3/e\hbar$, at which the field energy is sufficient to create electron positron pairs from the vacuum, and calculated an effective Lagrangian that will replace the Maxwell Lagrangian at strong fields. In 1951, Schwinger
\cite{Schwinger1951,Schwinger1954A,Schwinger1954B}
gave an elegant quantum-field theoretic reformulation of their result in the spinor and scalar QED framework (see also 
\cite{Nikishov1969,Batalin1970}). The description was further extended to space-time dependent electromagnetic fields in Refs.~
\cite{Popov1971,Popov1972,Popov2001,Narozhnyi1970,Schubert2001,Dunne2005B,Kleinert2013}. 
The monographs
\cite{Itzykson2006,Kleinert2008,Greiner1985,Grib1980,Fradkin1991}
and the recent review articles \cite{Dunne2000,Dunne2005,Ruffini2010} can be consulted for more detailed calculations, discussions and bibliographies.
Since then, the properties of QED in strong
electromagnetic fields have become a
vast arena of theoretical research, awaiting experimental verification as well as
further theoretical understanding.

An interesting aspect of effective field theories in the strong-field
limit has recently been emphasized
in a completely different class of quantum field theories. These have the
 property of
developing  an anomalous power behavior in the strong-field limit.
It is experimentally observable  at the critical point in second-order
phase transitions, and for this reason
such a power behavior is also called \textit{critical behavior}. It
arises if
the so-called beta function
(also called the  Stueckelberg--Petermann
 function or the Gell-Mann--Low function)\cite{Stueckelberg1953,GellMann1954},
which governs the logarithmic growth of the coupling strength for varying
energy scale,
has a fixed point in the infrared. In such theories, it is possible to
take the theory to the limit of infinite coupling strength.
The effective action can usually be  calculated in perturbation theory
as a power series
in the fields. The coefficients are the one-particle irreducible
$n$-point vertex functions
of the  theory.  In the limit of large field strength, this power series
  can be shown to develop
an anomalous power behavior with irrational exponents \cite{Kleinert2001Book,Kleinert403}.
Also the gradient terms in this effective action show anomalous powers
\cite{TalkTaipei}.
 
In the past many authors have  argued that in QED such a fixed point may exist
\cite{Johnson1963,Johnson1964,Maris1964,Maris1965, Frishman1965}
and could ultimately explain the numerical value of the fine structure constant. However it is presently believed to be absent. Lattice simulations as well as functional renormalization group methods show that chiral symmetry breaking prevents QED from reaching a fixed point  \cite{Gockeler1998, Gies2004}. This is supported by studies of the Gell-Mann--Low function \cite{Suslov2001}. However a fixed point might exist if one adds gravity to the theory \cite{Harst2011}.
In this article we shall not assume the existence of such a fixed point, but
point out that at strong fields, the effective action in the weak-coupling expansion, namely the Euler-Heisenberg effective action,  exhibits a
power behavior that is typical
for critical phenomena.

This anomalous power behavior in the weak-coupling expansion can be seen as a first step in the direction of a strong-coupling QED theory, by analogy with the above mentioned behavior for strong-coupling quantum field theories. This could be reached by using a technique to go from diverging weak-coupling series to a converging strong-coupling series which was developed in the context of \(\phi^4\)-theories \cite{Janke1995,Kleinert403,Kleinert2001Book}. However these calculations are left for future work. 

In QED, the nonperturbative, one-loop effective action takes the form
\begin{align}
 \Delta \mathcal{L}_\text{eff}[A]=-\i\, \text{tr} \log\left[\frac{\i\slashed{\partial}-e\slashed{A}(x)-m_e}{\i\slashed{\partial}-m_e}\right], \label{eq:1}
\end{align}
with the classical external gauge potential \(A_\mu\) and Feynman's slashed notation \(\slashed{v}:=\gamma^\mu v_{\mu}\).
Heisenberg and Euler  \cite{Heisenberg1936} and Weisskopf \cite{Weisskopf1936} showed that for a constant classical external field this can be brought to the form
\begin{align}
 \Delta \mathcal{L}_\text{eff}=-\frac{1}{2(2\pi)^2}\int_0^\infty\frac{ds}{s}\left[e^2\epsilon\beta\coth(s e \epsilon)\cot(s e \beta)-\frac{1}{s^2 }-\frac{e^2}{3}\left(\epsilon^2-\beta^2\right)\right]\text{e}^{-\i s (m_e^2-\i\eta)},
\label{eq:EH}
\end{align}
where \(\epsilon\) and \(\beta\) are Lorentz invariant variables defined by
\begin{align}
 \epsilon^2-\beta^2:=\vec{E}^2-\vec{B}^2:=-\frac{1}{2}F_{\mu\nu}F^{\mu\nu}=:2S,\hspace{1cm} \epsilon\beta:=\vec{E}\cdot\vec{B}:=-F_{\mu\nu}\tilde{F}^{\mu\nu}=:P
\end{align}
where \(\vec{E}\) and \(\vec{B}\) are the electric and magnetic field strength, \(F_{\mu\nu}=\partial_\mu A_\nu-\partial_\nu A_\mu\) is the field strength tensor and \(S \und P\) are scalar and pseudo-scalar combinations thereof. Explicitly, the quantities \(\epsilon\) and \(\beta\) are  given by
\begin{align}
\begin{split}
\epsilon=&\sqrt{\sqrt{S^2+P^2}+S},\\
\beta=&\sqrt{\sqrt{S^2+P^2}-S}. \label{eq:epsilonbeta}
\end{split}
\end{align}
For scalar QED, the corresponding quantity reads
\begin{align}
 \Delta \mathcal{L}_\text{eff}[A]=\frac{i}{2}\, \text{tr} \log\left\{\frac{[\i\partial_\mu-eA_\mu(x)]^2-m_e^2}{-\partial_\mu^2-m_e^2}\right\},
\end{align}
 which for constant classical external fields can be brought in the form  \cite{Weisskopf1936}
\begin{align}
 \delta \mathcal{L}_\text{eff}=\frac{1}{4(2\pi)^2}\int_0^\infty\frac{ds}{s}\left[e^2\epsilon\beta\text{csch}(s e \epsilon)\csc(s e \beta)-\frac{1}{s^2 }+\frac{e^2}{6}\left(\epsilon^2-\beta^2\right)\right]\text{e}^{-\i s (m_e^2-\i\eta)},
\label{eq:EHscalar}
\end{align}
where \(\csc(x)=1/\sin(x)\) and \(\csch(x)=1/\sinh(x)\).
There are several analytic reformulations and studies of the Euler-Heisenberg Lagrangians (\ref{eq:EH}) and (\ref{eq:EHscalar}) in Refs.~\cite{Heyl1996,Mielniczuk1982,Valluri1993,Cho2001,Ruffini2006,Kleinert}.

In Ref.~\cite{Weisskopf1936}, Weisskopf showed that the leading behavior of the effective Lagrangian (\ref{eq:EHscalar}) for strong magnetic and electric fields is logarithmic for vanishing electric and magnetic fields respectively. In this article we generalize this for several special cases of the variables \(P,S\) and \(\epsilon,\beta\), and express the results compactly as a fractional generalization of the Lagrangians (\ref{eq:EH}) and (\ref{eq:EHscalar})
\begin{align}
\mathcal{L}_\text{eff}=\frac{1}{2}E_c^{-2\delta}(\vec{E}^2-\vec{B}^2)\left({|\vec{E}^2-\vec{B}^2|}
{|\vec{E}\cdot\vec{B}|}\right)^{\delta/2}+\ldots,
\label{result:final}
\end{align}
which is valid in the limit of strong electromagnetic fields, with the anomalous power $\delta=e^2/(12\pi)$ for spinor QED and $\delta_S=\delta/4$ for scalar QED.

In Section \ref{sec:sumboth}, we briefly review the series representation of the Euler-Heisenberg-Lagrangian (\ref{eq:EH}) and (\ref{eq:EHscalar}), which has been derived in various places over the years \cite{Mielniczuk1982,Valluri1993,Cho2001,Ruffini2006}. In Section \ref{sec:strongfield}, we derive the fractional expression (\ref{result:final}) by using this series representation and performing a strong-field expansion up to the leading logarithmic order. In order to make the text and ideas most transparent, we relegate detailed technical calculations to the Appendices.

\section{Reformulation of the Euler-Heisenberg Lagrangian}
\label{sec:sumboth}
In this Section we use an identity to bring the Euler-Heisenberg Lagrangian (\ref{eq:EH}) to a form that will be used in section \ref{sec:strongfield} to go to strong fields. This reformulation has been done similarly in the past \cite{Mielniczuk1982,Valluri1993,Cho2001,Ruffini2010} for spinor QED. We shortly review this in Section \ref{sec:sum}. For scalar QED the procedure is presented in Section \ref{sec:sumscalar}.

\subsection{Reformulation of the spinor Euler-Heisenberg Lagrangian}
\label{sec:sum}
A useful expansion, to be derived in Appendix \ref{app:sum}, is \cite{Berndt1988}:
\begin{align}
\begin{split}
 e^2\epsilon\beta\coth(s e \epsilon)\cot(s e \beta)-\frac{1}{s^2}-\frac{e^2}{3}(\epsilon^2-\beta^2)=&2s^2\sum_{m=1}^\infty\frac{e \epsilon}{ s_m}\frac{\coth(e\epsilon s_m)}{s^2- s_m^2}
-2s^2\sum_{n=1}^{\infty} \frac{e \beta}{ s_n}\frac{\coth(e\beta s_n)}{s^2+ s_n^2}\label{eq:appsum}
\end{split}
\end{align}
with \(s_n=n\pi/(e \epsilon)\) and \(s_m=m\pi/(e \beta)\). Inserting that into (\ref{eq:EH}), it becomes
\begin{align}
 \Delta \mathcal{L}_\text{eff}=-\frac{1}{(2\pi)^2}\int_0^\infty ds\,s\left[\sum_{m=1}^\infty\frac{e \epsilon}{ s_m}\frac{\coth(e\epsilon s_m)}{s^2- s_m^2}-\sum_{n=1}^{\infty} \frac{e \beta}{ s_n}\frac{\coth(e\beta s_n)}{s^2+ s_n^2}\right]\text{e}^{-\i s (m_e^2-\i\eta)}.
\end{align}
We now rotate the integration contour to \(s=i \tau/(m_e^2 -\i \eta) \) where \(\tau\) runs along the real axis from zero to infinity, and the mass carries the usual negative infinitesimal part \(-\i\eta\) to ensure that the electron wave function goes to zero at infinite time, and we obtain
\begin{align}
 \Delta \mathcal{L}_\text{eff}=-\frac{m_e^2}{(2\pi)^2}\int_0^\infty d\tau\,\tau\left[
\sum_{m=1}^\infty\frac{e \epsilon}{\sigma_m}\frac{\coth(e\epsilon\tau_m)}{\tau^2-(\i\sigma_m)^2}
-\sum_{n=1}^{\infty} \frac{e \beta}{\sigma_n}\frac{\coth(e\beta\tau_n)}{\tau^2-\sigma_n^2}
\right]\text{e}^{-\tau},
\end{align}
with \(\sigma_k=s_k (m_e^2-\i\eta)\) for \(k=m,n\). Performing the integrals over \(\tau\), we arrive at the effective Lagrangian of QED. It has an imaginary part from the poles at \(\i\sigma_m \und \sigma_n\) in the integral, leading to the so-called Schwinger decay rate of the vacuum \cite{Schwinger1951,Schwinger1954A,Schwinger1954B}
\begin{align}
 \Delta \mathcal{L}_\text{eff}^\Im=\frac{\i\,m_e^2}{4\pi}\sum_{n=1}^{\infty} \frac{e \beta}{\sigma_n}{\coth\left(\frac{e\beta}{m_e^2}\tau_n\right)}\text{e}^{-\sigma_n} \label{eq:sumIm}.
\end{align}
The real part \(\Delta \mathcal{L}_\text{eff}^\Re\) can be derived with the help of the principal value integral (see \cite{Grad2000}, Eq.~3.354)
\begin{align}
 \mathcal{P}\int_0^\infty ds \frac{s e^{-s}}{s^2-z^2}=-\frac{1}{2}J(z):=-\frac{1}{2}\left({\text{e}^{-z}\text{Ei}(z)+\text{e}^z\text{Ei}(-z)}\right), \label{eq:J}
\end{align}
where \(\Ei(x)\) denotes the Exponential integral (see \cite{Grad2000}, Eq.~8.211)
\begin{align}
\Ei(x)=-\int_{-x}^\infty \frac{\text{e}^-t}{t}\,dt.
\end{align}
The result  is
\begin{align}
 \Delta \mathcal{L}_\text{eff}^\Re=\frac{m_e^2}{2(2\pi)^2}
\left[ 
\sum_{m=1}^{\infty} 
\frac{e \epsilon}{\sigma_m} 
\coth\left(\frac{e\epsilon}{m_e^2}\sigma_m\right) 
J(\i\sigma_m ) 
-
\sum_{n=1}^{\infty} 
\frac{e \beta}{\tau_n} 
\coth\left(\frac{e\beta}{m_e^2}\sigma_n\right) 
J(\sigma_n)
\right]. \label{eq:sumRe}
\end{align}
Equation (\ref{eq:sumRe}) was first presented in Ref. \cite{Mielniczuk1982}, later in different versions in \cite{Valluri1993,Cho2001}. The above version does not quite conform with standard mathematical notation, since the Exponential integral is usually defined for real values only (for details see \cite{Jentschura2002}). Nevertheless, since it is easily extended to purely imaginary values, using the sine and cosine integral (see Appendix \ref{app:sum2} 
for details) we proceed with the slightly improper notation for convenience.

Combining (\ref{eq:sumIm}) and (\ref{eq:sumRe}) and re-substituting \(\sigma_n\) and \(\sigma_m\) we find the total one-loop effective Lagrangian of QED for constant fields
\begin{align}
\begin{split}
\Delta \mathcal{L}_\text{eff}&=\frac{e^2}{(2\pi)^3}\sum_{n=1}^\infty \frac{\beta \epsilon}{ n}\left[
\coth\left(\pi \frac{\epsilon}{\beta}n\right)J\left(\frac{\i n \pi E_c}{\beta}\right)
-
\coth\left(\pi \frac{\beta}{\epsilon}n\right)\left[J\left(\frac{n \pi E_c}{\epsilon}\right)-2\pi \i\exp\left(-\frac{n \pi E_c}{\epsilon}\right)\right]
\right]. \label{eq:full}
\end{split}
\end{align}
Observe that this  is not invariant under the duality transformation \(\epsilon\rightarrow \i\beta, \beta\rightarrow -\i \epsilon\). This is due do the fact that for the extraction of the imaginary part we assumed \(\epsilon\text{ and } \beta\) to be real. There exists however  the possibility to incorporate the imaginary part into a formulation via analytic continuation using incomplete Gamma functions. This and its significance for the duality has been studied at length in the literature \cite{Jentschura2002,Bae2000}. We are however only interested in the real part, which also in the above formulation is invariant under the duality transformation.

\subsection{Reformulation of the scalar Euler-Heisenberg Lagrangian}
\label{sec:sumscalar}
 We now reformulate the scalar Euler-Heisenberg Lagrangian in the same way. Here (\ref{eq:appsum}) becomes (see Appendix \ref{app:scalaridentity})\cite{Berndt1988}
\begin{align}
\begin{split}
 e^2\epsilon\beta\text{csch}(s e \epsilon)\csc(s e \beta)-\frac{1}{s^2}+\frac{e^2}{6}(\epsilon^2-\beta^2)=&2s^2\sum_{m=1}^\infty(-1)^m\frac{e \epsilon}{ s_m}\frac{\text{csch}(e\epsilon s_m)}{s^2- s_m^2}
-2s^2\sum_{n=1}^{\infty} (-1)^n\frac{e \beta}{ s_n}\frac{\text{csch}(e\beta s_n)}{s^2+ s_n^2},
\end{split}
\end{align}
and (\ref{eq:sumIm}) changes slightly to
\begin{align}
 \Delta \mathcal{L}_\text{eff}^\Im=-\frac{\i m_e^2}{8\pi}\sum_{n=1}^{\infty}(-1)^n \frac{e \beta}{\tau_n}{\csch\left(\frac{e\beta}{m_e^2}\sigma_n\right)}\text{e}^{-\sigma_n},
\end{align}
a form first found by Popov \cite{Popov1972,Dunne2005}. Using the principal value integral (\ref{eq:J}),  the real part of the scalar effective Lagrangian becomes
\begin{align}
 \Delta \mathcal{L}_\text{eff}^\Re=\frac{m_e^2}{4(2\pi)^2}
\left[ 
\sum_{n=1}^{\infty}(-1)^n
\frac{e \beta}{\sigma_n} 
\csch\left(\frac{e\beta}{m_e^2}\sigma_n\right) 
J(\sigma_n)
-\sum_{m=1}^{\infty}(-1)^m
\frac{e \epsilon}{\sigma_m} 
\csch\left(\frac{e\epsilon}{m_e^2}\sigma_m\right) 
J(\i\sigma_m m_e^2) \label{eq:scalarEHRe}
\right],
\end{align}
and the full scalar Lagrangian reads, by analogy with (\ref{eq:full})
\begin{align}
\begin{split}
\Delta \mathcal{L}_\text{eff}&=\frac{e^2}{2(2\pi)^3}\sum_{n=1}^\infty(-1)^n \frac{\beta \epsilon}{ n}\left[\csch\left(\pi \frac{\beta}{\epsilon}n\right)
\left[J\left(\frac{n \pi E_c}{\epsilon}\right)-2\pi \i\exp\left(-\frac{n \pi E_c}{\epsilon}\right)\right]-\csch\left(\pi \frac{\epsilon}{\beta}n\right)J\left(\frac{\i n \pi E_c}{\beta}\right)\right]. \label{eq:scalarfull}
\end{split}
\end{align}
As in the spinor case, the duality transformation \(\epsilon\rightarrow \i\beta, \beta\rightarrow -\i \epsilon\) does not hold for the whole Lagrangian, for the same reasons as in the spinor case, with the same cure.

\section{Strong field approximation}
\label{sec:strongfield}
We now use the formulations of Sections \ref{sec:sum} and \ref{sec:sumscalar}  to find the leading terms in the approximation of the strong field limit (\(|\vec{E}|,\,|\vec{B}|\gg E_c\)). Since Eqs.~(\ref{eq:full}) and (\ref{eq:scalarfull}) are expressed in terms of the Lorentz invariant terms \(S=\vec{E}^2-\vec{B}^2\) and \(P=\vec{E}\cdot\vec{B}\), we must ensure that this limit is properly reflected in these variables. There is no direct way to translate the above limit to the variables \(\epsilon,\,\beta\). Thus we shall look at some special cases:
\begin{enumerate}
\item \(|S/P|\gg1\) ,
\item  \(\epsilon,\,\beta\gg E_c\) and \(\epsilon/\beta\sim\mathcal{O}(1)\) ,
\item  \(|P/S|\gg1\).
\end{enumerate}
The first case, where the component of the magnetic field in the direction of the electric field is small,  is discussed in detail in  \cite{Heyl1996} for the spinor case. We briefly revisit this case in Section \ref{sec:Psmall}, and discuss the scalar case in Section \ref{sec:PsmallScalar}.
In Sections \ref{sec:Pbig} and \ref{sec:PbigScalar}, we study the second case for spinor and scalar QED, respectively, which is a generalization of the third case to be studied in Sections \ref{sec:Ssmall} and \ref{sec:SsmallScalar}.
In six cases we find that the leading order correction is logarithmic, giving the possibility to rewrite the result as a power with an anomalous exponent. We additionally find that the only difference between scalar and spinor QED lies in a factor of 4 for this exponent.

\subsection{Strong field approximation for spinor QED}
\subsubsection{Small-\texorpdfstring{\(P\)}{P} expansion}
\label{sec:Psmall}
As described in Reference \cite{Heyl1996} for the case that the component of the magnetic field in direction of the electric field is small we can expand the corrections to the Lagrangian in the following way
\begin{align}
 \Delta \mathcal{L}(S,P)=\Delta \mathcal{L}(S,0)+P\left.\frac{\partial \Delta \mathcal{L}(S,P)}{\partial P}\right|_{P=0}+\ldots\,. \label{eq:SmallPExp}
\end{align}
Here we concentrate on the lowest order which means \(P=0\). Looking at (\ref{eq:epsilonbeta}), we see that this corresponds to \(\epsilon=\sqrt{2S},\,\beta=0 \) for \(S>0\) and \(\epsilon=0,\,\beta=\sqrt{-2S},\) for \(S<0\). These cases have been studied in \cite{Ruffini2010,Kleinert}. Using \(\lim_{z\rightarrow\infty}J(\i z)=0\)\text{, as well as } \(\lim_{z\rightarrow 0} z\,\coth(a z)=1/a\) we find the real part of the Lagrangian for \(\beta=0\)
\begin{align}
\Delta \mathcal{L}_{\beta=0}^\Re&=-\frac{e^2 \epsilon^2}{8 \pi^4}\sum_{n=1}^\infty \frac{1}{ n^2}J\left(\frac{n \pi E_c}{\epsilon}\right).
\end{align}
We can now look at what happens in the limit of strong fields \(\sqrt{2|S|}\gg E_c\). Then we can expand the exponential and exponential integral of \(J(x)\) defined in (\ref{eq:J}) for small \(x\).

It is important that we first perform the summation over \(n\) and then sort the orders of the strong field \(\epsilon\). The summation formulas needed for that can be found in Appendix \ref{app:sum2}. Using (\ref{eq:nsquaresumfull}) for real or purely imaginary \(\epsilon\) we find
\begin{align}
\Delta \mathcal{L}_{\beta=0}^\Re&=-\frac{e^2 \epsilon^2}{24 \pi^2}\log\left(\frac{\pi E_c}{\left|\epsilon\right|}\right)
- \frac{e^2 \epsilon^2}{8 \pi^4}\left(-2\zeta'(2)+\gamma \frac{\pi^2}{3}\right)+\ldots, \label{eq:vf1}
\end{align}
whereas for \(\epsilon=0\), the result is
\begin{align}
\Delta \mathcal{L}_{\epsilon=0}^\Re&=\frac{e^2 \beta^2}{24 \pi^2}\log\left(\frac{\pi E_c}{\left|\beta\right|}\right)
+ \frac{e^2 \epsilon^2}{8 \pi^4}\left(-2\zeta'(2)+\gamma \frac{\pi^2}{3}\right)+\ldots\,. \label{eq:vf2}
\end{align}
The first terms are those derived by Weisskopf \cite{Weisskopf1936} for vanishing magnetic (\(\epsilon=E\,,\beta=0\)) or electric (\(\epsilon=0\,,\beta=B\)) fields respectively (see also \cite{Kleinert}, Eq.~(13.435) and \cite{Ruffini2010}, Eq.~(199)). Here \(\zeta'(z)\) is the derivative of the Riemann zeta function.

Using \(\epsilon=\sqrt{2S}\) or \(\beta=\sqrt{-2S}\) we can now write 
\begin{align}
\mathcal{L}^{\Re}=&\frac{1}{2}(\vec{E}^2-\vec{B}^2)+\frac{e^2 }{24\pi^2}(\vec{E}^2-\vec{B}^2)\log\left(\frac{|\vec{E}^2-\vec{B}^2|}{E_c^2}\right)+\mathcal{O}\left(\frac{S}{E_c^2},\frac{P}{S}\right).
\end{align}
This can be brought into the form
\begin{align}
\mathcal{L}^{\Re}=&\frac{1}{2}E_c^{-\delta}(\vec{E}^2-\vec{B}^2){|\vec{E}^2-\vec{B}^2|}^{\delta/2}+\ldots, \label{eq:resultPsmall}
\end{align}
where we define the anomalous power 
\begin{align}
\delta:=\frac{e^2}{12 \pi}.
\end{align}
This is the same result as proposed in \cite{Kleinert}, Eq.~(13.436).
\subsubsection{General strong field case \texorpdfstring{(\(\beta,\,\epsilon\gg E_c\))}{b,e>>Ec}}
\label{sec:Pbig}
If we want to know more about the case \(\beta\ne 0,\, \epsilon\ne0\) in strong fields, we start from (\ref{eq:full}) and split the real part of the action into two parts
\begin{align}
&\Delta \mathcal{L}^\Re=\Delta\mathcal{L}^\epsilon+\Delta\mathcal{L}^\beta,
\end{align}
where
\begin{align}
\Delta \mathcal{L}^\epsilon&=-\frac{e^2 \beta \epsilon}{8\pi^3}\sum_{n=1}^\infty \frac{1}{ n}\coth\left(\pi  n z\right)J\left(n x\right)
,&&& \Delta\mathcal{L}^\beta=\frac{e^2 \beta \epsilon}{8\pi^3}\sum_{n=1}^\infty \frac{1}{ n}\coth\left(\pi n /z \right)J\left(\i n  x/z\right),
\end{align}
with \(z= \beta/\epsilon\) and \(x= \pi E_c/\epsilon\). We expand this for \(\epsilon\gg1\), which means around \(x=0\). For this we must first perform the sum over \(n\) . 
Note that \(z\) is not necessary small. We shall treat it as a quantity of order 1 in our expansion, which implies that \(\epsilon\) and \(\beta\) are of the same order of magnitude. 

The calculations necessary for the sums are summarized in Appendix \ref{app:sum2}. Using  (\ref{eq:cothsumlog}) we find the logarithmic corrections to the Lagrangian for real or purely imaginary \(\epsilon\) and \(\beta\) to be 
\begin{align}
\Delta \mathcal{L}^{\epsilon}_{\log}= \frac{e^2 \beta \epsilon}{8\pi^3}\log(|x|)\left[\text{sgn}(z) \left(4\log\left(\eta(\i |z|)\right)+ \log(|x|)\right) +\frac{1}{3}\pi z\right], \label{eq:Lepsilon}
\end{align}
where 
\(\eta(z)\) is the Dedekind Eta function \cite{Siegel1954}. We can now use its property 
\begin{align}
 \eta\left(-\frac{1}{\tau}\right)=\eta(\tau)\sqrt{-\i\tau}, \label{eq:etaidentity}
\end{align}
rewritten as
\begin{align}
\log(\eta(\i z))=\frac{1}{2}\log\left(\eta\left(\frac{\i}{z}\right)\eta(\i z)\right)-\frac{1}{4}\log(z). \label{eq:logfomula}
\end{align}
after which (\ref{eq:Lepsilon}) becomes

\begin{align}
\Delta \mathcal{L}^{\epsilon}_{\log}= \frac{e^2\epsilon\beta}{8\pi^3}\log\left(\frac{\pi E_c}{|\epsilon|}\right)\,\text{sgn}\left(\frac{\beta}{\epsilon}\right)
\left[2\log\left(\eta\left(\i\left|\frac{\beta}{\epsilon}\right|\right)\eta\left(\i \left|\frac{\epsilon}{\beta}\right|\right)\right) +\log\left(\frac{\pi E_c}{|\beta|}\right) +\left|\frac{\beta}{\epsilon}\right| \frac{ \pi}{3}\right],
\end{align}
By analogy we find for the second part (observe that we can get this result directly from the above via the duality transformation \(\epsilon\rightarrow \i \beta ,\, \beta\rightarrow-\i\epsilon\))
\begin{align}
\Delta \mathcal{L}^{\beta}_{\log}=- \frac{e^2\epsilon\beta}{8\pi^3}\log\left(\frac{\pi E_c}{|\beta|}\right)\,\text{sgn}\left(\frac{\epsilon}{\beta}\right)
\left[2\log\left(\eta\left(\i\left|\frac{\beta}{\epsilon}\right|\right)\eta\left(\i \left|\frac{\epsilon}{\beta}\right|\right)\right)+ \log\left(\frac{\pi E_c}{|\epsilon|}\right) +\left|\frac{\epsilon}{\beta} \right|\frac{ \pi}{3}\right].
\end{align}
So that in the end we have
\begin{align}
\begin{split}
\Delta \mathcal{L}_{\log}^\Re =&- \frac{e^2}{24\pi^2}\left[ \epsilon^2\log\left(\frac{\pi E_c}{|\beta|}\right)-\beta^2\log \left(\frac{\pi E_c}{|\epsilon|}\right)  \right]+\text{sgn}\left(\frac{\epsilon}{\beta}\right)\frac{e^2 \beta \epsilon}{4\pi^3} \log\left(\frac{|\beta|}{|\epsilon|}\right)\log\left(\eta\left(\i\left|\frac{\beta}{\epsilon}\right|\right)\eta\left(\i \left|\frac{\epsilon}{\beta}\right|\right)\right) \label{eq:logepsilonbeta},
\end{split}
\end{align}
where the last term is only a function of the ratio of \(\epsilon\) and \(\beta\), and thus not of logarithmic order. Thus the result  reads
\begin{align}
 \Delta \mathcal{L}_{\log}^\Re =& \frac{e^2}{24\pi^2}\left[ \epsilon^2\log\left(\frac{|\beta|}{\pi E_c}\right)-\beta^2\log \left(\frac{|\epsilon|}{\pi E_c}\right)  \right].
\end{align}
For the total effective Lagrangian we find
\begin{align}
 \mathcal{L}^{\Re}=\frac{1}{2}(\epsilon^2-\beta^2)+\frac{e^2}{24\pi^2}\left[ \epsilon^2\log\left(\frac{|\beta|}{ E_c}\right)-\beta^2\log \left(\frac{|\epsilon|}{ E_c}\right)\right]+\mathcal{O}\left(\frac{\epsilon^2}{E_c^2},\frac{\beta^2}{E_c^2}\right).
\end{align}
This is formulated with an anomalous power as
\begin{align}
 \mathcal{L}^{\Re}=\frac{1}{2} E_c ^{-\delta}\left(\epsilon^{2}|\beta|^{\delta}-\beta^{2}|\epsilon|^{\delta}\right)+\ldots, \label{eq:resultepsilonbeta}
\end{align}
where the coefficient \(\delta=e^2/12 \pi\)
is the same as in the small-\(P\) case.
\subsubsection{Small-\texorpdfstring{\(S\)}{S} expansion}
\label{sec:Ssmall}
In the above section we studied the case of \(\epsilon, \beta\gg E_c\) while \(\epsilon/\beta\sim\mathcal{O}(1)\). While the first restriction means that \(|P|\gg E_c^2\) the second does not necessarily mean that \(S\) is small. To study the \(|P/S|\gg 1\) case we expand 
\begin{align}
\begin{split}
 \epsilon&=\sqrt{|P|}\left(1+\frac{1}{2}\frac{S}{|P|}\right)+\mathcal{O}\left(\frac{S^2}{P^2}\right),\\
 \beta&=\sqrt{|P|}\left(1-\frac{1}{2}\frac{S}{|P|}\right)+\mathcal{O}\left(\frac{S^2}{P^2}\right).
\end{split} \label{eq:epsbetaexpand}
\end{align}
If we insert this in the logarithmic order for the \(\epsilon,\beta\gg E_c\) case (\ref{eq:logepsilonbeta}) we find
\begin{align}
 \Delta \mathcal{L}_{\log}^\Re =&-\frac{e^2}{12 \pi^2} S \log\left(\frac{\pi E_c}{\sqrt{|P|}}\right)-\frac{e^2}{24 \pi^2 }S-\frac{e^2}{4\pi^3}S \log\left(\eta(\i)\right)+\mathcal{O}\left(\frac{S^2}{P^2}\right).
\end{align}
The last two terms are not of logarithmic growth so that we by discarding them find
\begin{align}
 \Delta \mathcal{L}_{\log}^\Re =&\frac{e^2}{24 \pi^2} S \log\left(\frac{|P|}{E_c^2}\right)+\mathcal{O}\left(\frac{S^2}{P^2}\right). \label{eq:smallSlog}
\end{align}
For the Lagrangian we find
\begin{align}
 \mathcal{L}^{\Re}=\frac{1}{2}(\vec{E}^2-\vec{B}^2)+\frac{e^2}{48 \pi^2} (\vec{E}^2-\vec{B}^2) \log\left(\frac{|\vec{E}\cdot\vec{B}|}{E_c^2}\right)+\mathcal{O}\left(\frac{S}{E_c^2},\frac{S^2}{P^2}\right) ,
\end{align}
which can be brought in the form
\begin{align}
\mathcal{L}^{\Re}=\frac{1}{2}E_c^{-\delta}(\vec{E}^2-\vec{B}^2){|\vec{E}\cdot\vec{B}|}^{\delta/2}+\ldots \label{eq:resultSsmall}
\end{align}
with the anomalous power \(\delta=e^2/12 \pi\).

\subsection{Strong field approximation for scalar QED}
\subsubsection{Small-\texorpdfstring{\(P\)}{P} expansion}
\label{sec:PsmallScalar}
As in section \ref{sec:Psmall} we concentrate on the first order of the expansion of the Lagrangian in \(P\) (\ref{eq:SmallPExp}) which is  related to \(P=0\) or \(\epsilon=\sqrt{2S},\,\beta=0 \) for \(S>0\) and \(\epsilon=0,\,\beta=\sqrt{-2S},\) for \(S<0\). Using \(\lim_{z\rightarrow\infty}J(\i z)=0\)\text{ as well as } \(\lim_{z\rightarrow 0} z\,\csch(a z)=1/a\) we find the real part of the scalar Lagrangian for \(\beta=0\)  
\begin{align}
\Delta \mathcal{L}_{\beta=0}^\Re&=\frac{e^2 \epsilon^2}{16 \pi^4}\sum_{n=1}^\infty \frac{(-1)^n}{ n^2}J\left(\frac{n \pi E_c}{\epsilon}\right).
\end{align}
Using (\ref{eq:nsquaresumscalar}) for real or purely imaginary \(\epsilon\) we find
\begin{align}
\Delta \mathcal{L}_{\beta=0}^\Re&=-\frac{e^2 \epsilon^2}{96 \pi^2}\log\left(\frac{\pi E_c}{\left|\epsilon\right|}\right)
+ \frac{e^2 \epsilon^2}{16 \pi^4}\left(\zeta'(2)+[\log(2)-\gamma] \frac{\pi^2}{6}\right)+\ldots. \label{eq:vf1scalar}
\end{align}
Analogously we obtain for \(\epsilon=0\)
\begin{align}
\Delta \mathcal{L}_{\epsilon=0}^\Re&=\frac{e^2 \beta^2}{96 \pi^2}\log\left(\frac{\pi E_c}{\left|\beta\right|}\right)
- \frac{e^2 \beta^2}{16 \pi^4}\left(\zeta'(2)+[\log(2)-\gamma] \frac{\pi^2}{6}\right)+\ldots. \label{eq:vf2scalar}
\end{align}
Using \(\epsilon=\sqrt{2S}\) or \(\beta=\sqrt{-2S}\) we can now write 
\begin{align}
\mathcal{L}^{\Re}=&\frac{1}{2}(\vec{E}^2-\vec{B}^2)+\frac{e^2 }{96\pi^2}(\vec{E}^2-\vec{B}^2)\log\left(\frac{|\vec{E}^2-\vec{B}^2|}{E_c^2}\right)+\mathcal{O}\left(\frac{S}{E_C^2},\frac{P}{S}\right), \label{eq:PsmallLog}
\end{align}
which can be brought into the form
\begin{align}
\mathcal{L}^{\Re}=&\frac{1}{2}E_c^{-\delta'}(\vec{E}^2-\vec{B}^2){|\vec{E}^2-\vec{B}^2|}^{\delta'/2}+\ldots, \label{eq:resultPsmallscalar}
\end{align}
where we defined the anomalous power 
\begin{align}
\delta':=\frac{e^2}{48 \pi}.
\end{align}
\subsubsection{General strong field case \texorpdfstring{(\(\beta,\,\epsilon\gg E_c\))}{b,e>>Ec}}
\label{sec:PbigScalar}
To study the case \(\epsilon,\,\beta\gg E_c\) and \(\epsilon/\beta\sim\mathcal{O}(1)\) we split up the real part of the action (\ref{eq:scalarfull})
\begin{align}
&\Delta \mathcal{L}^\Re=\Delta\mathcal{L}^\epsilon+\Delta\mathcal{L}^\beta,
\end{align}
where
\begin{align}
&\Delta \mathcal{L}^\epsilon=\frac{e^2 \beta \epsilon}{16\pi^3}\sum_{n=1}^\infty \frac{(-1)^n}{ n}\csch\left(\pi n z \right)J\left(n x\right)
,\, &&& \Delta\mathcal{L}^\beta=-\frac{e^2 \beta \epsilon}{16\pi^3}\sum_{n=1}^\infty \frac{(-1)^n}{ n}\csch\left(\pi n/ z \right)J\left(\i n x/z\right),
\end{align}
with \(z= \beta/\epsilon\) and \(x= \pi E_c/\epsilon\). Now using (\ref{eq:cschsumlog})  we find
\begin{align}
\Delta \mathcal{L}^{\epsilon}_{\log}= \frac{e^2 \beta \epsilon}{16\pi^3}\log(|x|)\left[\frac{1}{3}\pi z-4 \,\text{sgn}(z) \log\left(\kappa(\i |z|)\right)  \right].
\end{align}
Where the function \(\kappa(x)\), defined in equation (\ref{eq:kappa}), unlike the Dedekind Eta function used in the spinor case in section \ref{sec:Pbig} is not a function which can be found in the literature. As shown in Appendix \ref{app:kappa} in Eq.~(\ref{eq:kappaidentity}) it is however possible to find a property for this function which is analog to (\ref{eq:etaidentity})  for the Dedekind Eta function. This identity  reads 
\begin{align}
 \kappa\left(-\frac {1}{\tau}\right)=\kappa(\tau) 
\end{align}
and can be brought into the form 
\begin{align}
\log(\kappa(\i z))=\frac{1}{2}\log\left(\kappa\left(\frac{\i}{z}\right)\kappa(\i z)\right).
\end{align}
We thus  by re-substituting \(x\) and \(z\) find
\begin{align}
\Delta \mathcal{L}^{\epsilon}_{\log}= \frac{e^2\epsilon\beta}{16\pi^3}\log\left(\frac{\pi E_c}{|\epsilon|}\right)\,\text{sgn}\left(\frac{\beta}{\epsilon}\right)
\left[ \left|\frac{\beta}{\epsilon}\right| \frac{ \pi}{6}-2\log\left(\kappa\left(\i\left|\frac{\beta}{\epsilon}\right|\right)\kappa\left(\i \left|\frac{\epsilon}{\beta}\right|\right)\right) \right]
\end{align}
and analogously
\begin{align}
\Delta \mathcal{L}^{\beta}_{\log}= -\frac{e^2\epsilon\beta}{16\pi^3}\log\left(\frac{\pi E_c}{|\epsilon|}\right)\,\text{sgn}\left(\frac{\epsilon}{\beta}\right)
\left[ \left|\frac{\epsilon}{\beta}\right| \frac{ \pi}{6}-2\log\left(\kappa\left(\i\left|\frac{\beta}{\epsilon}\right|\right)\kappa\left(\i \left|\frac{\epsilon}{\beta}\right|\right)\right) \right].
\end{align}
Adding these two equations we find
\begin{align}
\begin{split}
\Delta \mathcal{L}_{\log}^\Re =&- \frac{e^2}{96\pi^2}\left[ \epsilon^2\log\left(\frac{\pi E_c}{|\beta|}\right)-\beta^2\log \left(\frac{\pi E_c}{|\epsilon|}\right)  \right]-\text{sgn}\left(\frac{\epsilon}{\beta}\right)\frac{e^2 \beta \epsilon}{8\pi^3} \log\left(\frac{|\beta|}{|\epsilon|}\right)\log\left(\kappa\left(\i\left|\frac{\beta}{\epsilon}\right|\right)\kappa\left(\i \left|\frac{\epsilon}{\beta}\right|\right)\right).\label{eq:scalarlogepsilonbeta}
\end{split}
\end{align}
In analogy to (\ref{eq:logepsilonbeta}) the last term is not logarithmic since it only depends on the fraction of \(\epsilon \und \beta\). This means the logarithmic part takes the form
\begin{align}
\begin{split}
\Delta \mathcal{L}_{\log}^\Re =&- \frac{e^2}{96\pi^2}\left[ \epsilon^2\log\left(\frac{\pi E_c}{|\beta|}\right)-\beta^2\log \left(\frac{\pi E_c}{|\epsilon|}\right)  \right],
\end{split}
\end{align}
which leads to the Lagrangian
\begin{align}
 \mathcal{L}^{\Re}=\frac{1}{2}(\epsilon^2-\beta^2)+\frac{e^2}{96\pi^2}\left[ \epsilon^2\log\left(\frac{|\beta|}{ E_c}\right)-\beta^2\log \left(\frac{|\epsilon|}{ E_c}\right)\right]+\mathcal{O}\left(\frac{\epsilon^2}{E_c^2},\frac{\beta^2}{E_c^2}\right).
\end{align}
Or formulated with an anomalous power
\begin{align}
 \mathcal{L}^{\Re}=\frac{1}{2} E_c ^{-\delta'}\left(\epsilon^{2}|\beta|^{\delta'}-\beta^{2}|\epsilon|^{\delta'}\right)+\ldots, \label{eq:resultepsilonbetascalar}
\end{align}
where the coefficient \(\delta'=e^2/48 \pi\)
is the same as in the small \(P\) case.
\subsubsection{Small-\texorpdfstring{\(S\)}{S} expansion}
\label{sec:SsmallScalar}
We find the \(|P/S|\gg 1\) expansion from (\ref{eq:scalarlogepsilonbeta}) by expanding with the help of (\ref{eq:epsbetaexpand}) and find
\begin{align}
 \Delta \mathcal{L}_{\log}^\Re =&-\frac{e^2}{48 \pi^2} S \log\left(\frac{\pi E_c}{\sqrt{|P|}}\right)-\frac{e^2}{96 \pi^2 }S+\frac{e^2}{16\pi^3}S \log\left(\kappa(\i)\right)+\mathcal{O}\left(\frac{S^2}{P^2}\right).
\end{align}
By discarding the non logarithmic powers we find in analogy to (\ref{eq:smallSlog})
\begin{align}
 \Delta \mathcal{L}_{\log}^\Re =&\frac{e^2}{96 \pi^2} S \log\left(\frac{|P|}{E_c^2}\right)+\mathcal{O}\left(\frac{S^2}{P^2}\right). 
\end{align}
For the whole Lagrangian we find
\begin{align}
 \mathcal{L}^{\Re}=\frac{1}{2}(\vec{E}^2-\vec{B}^2)+\frac{e^2}{192 \pi^2} (\vec{E}^2-\vec{B}^2) \log\left(\frac{|\vec{E}\cdot\vec{B}|}{E_c^2}\right)+\mathcal{O}\left(\frac{S}{E_c^2},\frac{S^2}{P^2}\right) 
\end{align}
and thus we find
\begin{align}
\mathcal{L}^{\Re}=\frac{1}{2}E_c^{-\delta'}(\vec{E}^2-\vec{B}^2){|\vec{E}\cdot\vec{B}|}^{\delta'/2}+\ldots, \label{eq:resultSsmallscalar}
\end{align}
with the anomalous power \(\delta'=e^2/48 \pi\).

\section{Conclusions}

We conclude that in the strong-field expansion, the leading order behavior of the Euler-Heisenberg effective Lagrangian is logarithmic, and can be formulated as a power law for three different cases:
\begin{enumerate}
\item \(|S/P|\gg1\),
\item  \(\epsilon,\,\beta\gg E_c\) and \(\epsilon/\beta\sim\mathcal{O}(1)\),
\item  \(|P/S|\gg1\).
\end{enumerate}
The general form is the same for scalar and spinor QED. The only difference is a factor of four in the anomalous power $\delta$.

Let us mention here that the anomalous powers \(\delta\) and \(\delta_S\) are closely connected to the lowest Taylor coefficient of the expansion of the \(\beta\)-function
\begin{align}
 \beta(e)=\beta_1 e^3+\mathcal{O}( e^5),
\end{align}
where (see e.g.~\cite{Dunne2005})
\begin{align}
 \beta^\text{spinor}_1=\frac{1}{12\pi^2},~~~~~\beta^\text{scalar}_1=\frac{1}{48\pi^2},
\end{align}
 for spinor and scalar QED, respectively. This can be explained straightforwardly since there exists a well-known relation between the strong-field limit of the effective Lagrangian and the perturbative beta function which was first established in Ref.~\cite{Coleman1973} and reads (see e.g.~\cite{Dunne2005})
\begin{align}
 \Delta\mathcal{L}_\text{eff}\sim\frac{1}{4}(\beta_1e^2+\ldots)(\vec{E}^2-\vec{B}^2)\log(|\vec{E}^2-\vec{B}^2|).
\end{align}
By comparing with (\ref{eq:PsmallLog}) one sees that the anomalous powers are related to \(\beta_1\) by
\begin{align}
 \delta=\beta_1 e^2.
\end{align}

We have not been able to derive a result for \(S,P\gg E_c^2\). This case is equivalent to \(|\vec{E}|\gg|\vec{B}|\gg E_c\) or \(|\vec{B}|\gg|\vec{E}|\gg E_c\) while the fields are almost parallel.
If we combine the results (\ref{eq:resultPsmall}) and (\ref{eq:resultSsmall}) for the cases 1.~and 3.~we can conjecture the more general result Eq.~(\ref{result:final}). This correctly reduces to the cases 1.~and 3.~in the respective limits and thus is more general.
As a result, Eq.~(\ref{result:final}) defines a fractional formulation for QED in the regime of strong fields.  
Thus our finding exhibits an interesting similarity to the fractional quantum field theory discussed in Ref.~\cite{Kleinert2012}. 

As mentioned in the introduction this analogy might allow us to use the methods developed in \cite{Janke1995,Kleinert403,Kleinert2001Book} to get a converging strong-coupling expansion starting from the diverging weak-coupling series of QED. Thus the present paper presents a first step towards a strong-coupling theory for QED.

The Euler-Heisenberg Lagrangian is obtained in the configuration of constant electromagnetic fields. Nevertheless, for the case of smooth and slow variations of electromagnetic fields in space and time, it can be approximately used to study interesting effects like light-by-light scattering, photon splitting or electron-positron pair production (for reviews see \cite{Dunne2005,Ruffini2010}). This implies that the fractional QED obtained in this article could find some applications in the regime of strong electromagnetic fields, due to its elegant mathematical formulation. This is particularly important for the recent rapid developments of experimental facilities using novel strong laser sources to reach the field strength and intensity of theoretical interest. Such facilities include the Extreme Light Infrastructure (ELI)\footnote{see \url{http://www.extreme-light-infrastructure.eu/}}, the Exawatt Center for Extreme Light studies (XCELS)\footnote{see \url{http://www.xcels.iapras.ru/}}, or 
the High Power laser Energy Research (HiPER)\footnote{see \url{
http://www.hiper-laser.org/}} facility, which are planned to exceed powers of 100\,PW. Both theoretical and experimental studies of the QED of strong electromagnetical fields at the Sauter-Euler-Heisenberg scale $E_c$ promise to become increasingly  fascinating in the coming years.

\section*{Acknowledgements}

ES is supported by the Erasmus Mundus Joint Doctorate Program by Grant Number 2012-1710 from the EACEA of the European Commission.


\appendix

\section{ Elementary identities}

\subsection{Identity used for spinor QED}
\label{app:sum}
Here we derive the formula (\ref{eq:appsum}) used in Section \ref{sec:sum} to simplify the Euler-Heisenberg Lagrangian. 
We start from the well-known series representations for the cotangent and the hyperbolic cotangent function (see \cite{Grad2000}, Eq.~1.421)
\begin{align}
 \cot(\pi x)=&\frac{1}{\pi x}+\frac{2x}{\pi}\sum_{k=1}^{\infty}\frac{1}{x^2-k^2},\\
 \coth(\pi x)=&\frac{1}{\pi x}+\frac{2x}{\pi}\sum_{k=1}^{\infty}\frac{1}{x^2+k^2},
\end{align}
to find 
\begin{align}
 & e^2\epsilon\beta\coth(s e \epsilon)\cot(s e \beta)-\frac{1}{s^2}=4s^2\sum_{m,n=1}^{\infty}\frac{1}{s^2- s_m^2}\frac{1}{s^2+ s_n^2}
+2\sum_{m=1}^{\infty}\frac{1}{s^2- s_m^2}+2\sum_{n=1}^{\infty}\frac{1}{s^2+ s_n^2} \label{eq:neededsum},&
\end{align}
where we have introduced \( s_n=n\pi/(e \epsilon)\) and \( s_m=m\pi/(e \beta)\). The first sum can be decomposed as
\begin{align}
\sum_{m,n=1}^{\infty}\frac{1}{s^2- s_m^2}\frac{1}{s^2+ s_n^2}=&\sum_{m,n=1}^{\infty}\frac{1}{ s_m^2+ s_n^2}\left(\frac{1}{s^2- s_m^2}-\frac{1}{s^2+ s_n^2}\right). \label{eq:splitsum}
\end{align}
The individual sums in (\ref{eq:neededsum}) can be expressed in terms of the digamma function \(\psi(z)=\Gamma'(z)/\Gamma(z)\) as follows
\begin{align}
\begin{split}
 \sum_{m=1}^\infty \frac{1}{\left(\frac{\pi m}{e \beta}\right)^2+ s_n^2}=&\frac{e^2\beta^2}{\pi^2}\sum_{m=1}^\infty \frac{1}{m^2+\left(\frac{e\beta}{\pi} s_n\right)^2} \label{eq:suminbetween}\\
=&\frac{e \beta}{2 \pi  s_n}\i\left[\sum_{m=1}^\infty \frac{1}{m+ \i \frac{e\beta}{\pi} s_n}-\sum_{m=1}^\infty \frac{1}{m- \i \frac{e\beta}{\pi} s_n}\right]\\
=&\frac{e \beta}{2 \pi  s_n}\i\left[-\psi\left(1+\i\frac{e\beta}{\pi} s_n\right)+\psi\left(1-\i\frac{e\beta}{\pi} s_n\right)\right]
\end{split}
\end{align}
where we have used the series representation (see \cite{Grad2000}, Eq.~8.363)
\begin{align}
\psi(x)-\psi(y)=\sum_{k=0}^\infty\left(\frac{1}{k+y}-\frac{1}{k+x}\right). \label{eq:psiseries}
\end{align}
The result is simplified with the identity
\begin{align}
 \psi(1-\i x)-\psi(1+\i x)=-\i\left(\pi \coth(\pi x)-\frac{1}{x}\right), \label{eq:psi}
\end{align}
that can be derived from the reflection formula (see \cite{Grad2000}, Eq.~8.334) of the Gamma function
\begin{align}
\Gamma(1-z)\Gamma(1+z)=\Gamma(1-z)\Gamma(z)z=\frac{z \pi}{\sin(\pi z)}.
\end{align}
 The logarithmic derivative of this 
\begin{align}
 \psi(z+1)-\psi(1-z)=\frac{1}{z}-\pi \cot(\pi z) \label{eq:psisum}
\end{align}
and \(\cot(\i z)=-\i \coth(z)\) lead directly to (\ref{eq:psi}). This allows us to simplify the double sum in (\ref{eq:splitsum}). First we have from (\ref{eq:suminbetween})
\begin{align}
  \sum_{m=1}^\infty \frac{1}{ s_m^2+ s_n^2}=\frac{1}{2 }\left(\frac{e \beta}{ s_n}\coth(e\beta s_n)-\frac{1}{ s_n^2}\right),
\end{align}
and
\begin{align}
  \sum_{n=1}^\infty \frac{1}{ s_m^2+ s_n^2}=\frac{1}{2 }\left(\frac{e \epsilon}{ s_m}\coth(e\epsilon s_m)-\frac{1}{ s_m^2}\right).
\end{align}
Now we sum the first sum in equation (\ref{eq:splitsum}) over \(m\) and the second over \(n\), and insert everything in (\ref{eq:neededsum}) to get 
\begin{align}
\begin{split}
 e^2\epsilon\beta\coth(s e \epsilon)\cot(s e \beta)-\frac{1}{s^2}=&2s^2\sum_{m=1}^\infty\frac{e \epsilon}{ s_m}\coth(e\epsilon s_m)\frac{1}{s^2- s_m^2}
-2s^2\sum_{n=1}^{\infty} \frac{e \beta}{ s_n}\coth(e\beta s_n)\frac{1}{s^2+ s_n^2}
\\
&+2\sum_{m=1}^{\infty}\frac{1-\frac{s^2}{ s_m^2}}{s^2- s_m^2}+2\sum_{n=1}^{\infty}\frac{1+\frac{s^2}{ s_n^2}}{s^2+ s_n^2}.
\end{split}
\end{align}
The last two sums simplify to
\begin{align}
-2 \sum_{m=1}^\infty  s_m^{-2}+2 \sum_{n=1}^\infty  s_n^{-2}=\frac{2 e^2}{\pi^2}(\epsilon^2-\beta^2)\sum_{k=1}^\infty k^{-2}=\frac{e^2}{3}(\epsilon^2-\beta^2),
\end{align}
so that we arrive at (\ref{eq:appsum})
\begin{align}
\begin{split}
 e^2\epsilon\beta\coth(s e \epsilon)\cot(s e \beta)-\frac{1}{s^2}-\frac{e^2}{3}(\epsilon^2-\beta^2)=&2s^2\sum_{m=1}^\infty\frac{e \epsilon}{ s_m}\frac{\coth(e\epsilon s_m)}{s^2- s_m^2}
-2s^2\sum_{n=1}^{\infty} \frac{e \beta}{ s_n}\frac{\coth(e\beta s_n)}{s^2+ s_n^2}. \label{eq:aappsum}
\end{split}
\end{align}

\subsection{Identity used for scalar QED}
\label{app:scalaridentity}
Here we derive the analog of (\ref{eq:aappsum}) for the bosonic case. We start from series representations of the co-secant and hyperbolic co-secant function (see \cite{Grad2000}, Eq.~1.422)\footnote{The second one is straightforwardly obtained from the first one via \(\text{csch}(x)=\i \csc(\i x)\).}
\begin{align}
 \csc(\pi x)=&\frac{1}{\pi x}+\frac{2x}{\pi}\sum_{k=1}^{\infty}\frac{(-1)^k}{x^2-k^2},\\
 \text{csch}(\pi x)=&\frac{1}{\pi x}+\frac{2x}{\pi}\sum_{k=1}^{\infty}\frac{(-1)^k}{x^2+k^2}.
\end{align}
Combining these two yields
\begin{align}
\begin{split}
 e^2\epsilon\beta\text{csch}(s e \epsilon)\csc(s e \beta)-\frac{1}{s^2}=&4s^2\sum_{m,n=1}^{\infty}\frac{(-1)^m}{s^2- s_m^2}\frac{(-1)^n}{s^2+ s_n^2}\\
&+2\sum_{m=1}^{\infty}\frac{(-1)^m}{s^2- s_m^2}+2\sum_{n=1}^{\infty}\frac{(-1)^n}{s^2+ s_n^2} \label{eq:neededsumscalar},
\end{split}
\end{align}
where we introduced \( s_n=n\pi/(e \epsilon)\) and \( s_m=m\pi/(e \beta)\). The first sum can now be split as
\begin{align}
\sum_{m,n=1}^{\infty}\frac{(-1)^m}{s^2- s_m^2}\frac{(-1)^n}{s^2+ s_n^2}=&\sum_{m,n=1}^{\infty}\frac{(-1)^{n+m}}{ s_m^2+ s_n^2}\left(\frac{1}{s^2- s_m^2}-\frac{1}{s^2+ s_n^2}\right). \label{eq:splitsumscalar}
\end{align}
By analogy with (\ref{eq:suminbetween}) we find 
\begin{align}
\begin{split}
 \sum_{m=1}^\infty \frac{(-1)^m}{\left(\frac{\pi m}{e \beta}\right)^2+ s_n^2}=&\frac{e^2\beta^2}{\pi^2}\sum_{m=1}^\infty \frac{(-1)^m}{m^2+\left(\frac{e\beta}{\pi} s_n\right)^2}\\
=&\frac{e \beta}{2 \pi  s_n}\i\left[\sum_{m=1}^\infty \frac{(-1)^m}{m+ \i \frac{e\beta}{\pi} s_n}-\sum_{m=1}^\infty \frac{(-1)^m}{m- \i \frac{e\beta}{\pi} s_n}\right]\\
=&\frac{e \beta}{4 \pi  s_n}\i\left[-\psi\left(1+\i\frac{e\beta}{2\pi} s_n\right)+\psi\left(1-\i\frac{e\beta}{2\pi} s_n\right)\right.\\
&\hspace{1cm}\left. +\psi\left(\frac{1}{2}+\i\frac{e\beta}{2\pi} s_n\right)-\psi\left(\frac{1}{2}-\i\frac{e\beta}{2\pi} s_n\right) \right]
\end{split}
\end{align} 
where we again used (\ref{eq:psiseries}). Now we use (\ref{eq:psisum}) as well as (see \cite{Grad2000}, Eq.~8.366.9)
\begin{align}
 \psi\left(\frac{1}{2}+z\right)= \psi\left(\frac{1}{2}-z\right)+\pi\tan(\pi z)
\end{align}
and \(\tan(\i z)=\i\tanh(z)\) to find
\begin{align}
  \sum_{m=1}^\infty \frac{(-1)^m}{ s_m^2+ s_n^2}&=\frac{1}{2 }\left(\frac{e \beta}{2 s_n}\left[\coth\left(\frac{e\beta s_n}{2}\right)-\tanh\left(\frac{e\beta s_n}{2}\right)\right]-\frac{1}{ s_n^2}\right)\\
 &=\frac{1}{2 }\left(\frac{e \beta}{ s_n}\text{csch}(e\beta s_n)-\frac{1}{ s_n^2}\right) 
\end{align}
Summing the first sum in equation (\ref{eq:splitsumscalar}) over \(m\)  and the second over \(n\),  we get 
\begin{align}
\begin{split}
 e^2\epsilon\beta\text{csch}(s e \epsilon)\csc(s e \beta)-\frac{1}{s^2}=&2s^2\sum_{m=1}^\infty\frac{e \epsilon}{ s_m}\text{csch}(e\epsilon s_m)\frac{(-1)^m}{s^2- s_m^2}-2s^2\sum_{n=1}^{\infty} \frac{e \beta}{ s_n}\text{csch}(e\beta s_n)\frac{(-1)^n}{s^2+ s_n^2}\\
+&2\sum_{m=1}^{\infty}(-1)^m\frac{1-\frac{s^2}{ s_m^2}}{s^2- s_m^2}+2\sum_{n=1}^{\infty}(-1)^n\frac{1+\frac{s^2}{ s_n^2}}{s^2+ s_n^2}.
\end{split}
\end{align}
The two last sums are now combined to
\begin{align}
-2 \sum_{m=1}^\infty (-1)^m s_m^{-2}+2 \sum_{n=1}^\infty (-1)^n s_n^{-2}=\frac{2 e^2}{\pi^2}(\epsilon^2-\beta^2)\sum_{k=1}^\infty(-1)^k k^{-2}=-\frac{e^2}{6}(\epsilon^2-\beta^2),
\end{align}
and lead to
\begin{align}
\begin{split}
 e^2\epsilon\beta\text{csch}(s e \epsilon)\csc(s e \beta)-\frac{1}{s^2}+\frac{e^2}{6}(\epsilon^2-\beta^2)=&2s^2\sum_{m=1}^\infty(-1)^m\frac{e \epsilon}{ s_m}\frac{\text{csch}(e\epsilon s_m)}{s^2- s_m^2}-2s^2\sum_{n=1}^{\infty} (-1)^n\frac{e \beta}{ s_n}\frac{\text{csch}(e\beta s_n)}{s^2+ s_n^2}.
\end{split}
\end{align}

\section{Summation formulas}
\label{app:sum2}
For the calculations in the main part we use the series expansion of the exponential integral function for real and purely imaginary arguments\footnote{For real values see \cite{Grad2000}, Eq.~8.214. The series for purely imaginary values can be derived via \(\text{Ei}(\pm \i x)=\text{ci}(x)\pm \text{si}(x)\) and the respective series for the sine and cosine integral (see \cite{Grad2000}, Eqs.~8.232 and 8.233).\label{fn:Ei}}
\begin{align}
 \text{Ei}(z)=\gamma+\log^*(z)+\sum_{k=1}^\infty \frac{z^k}{k k!} \label{eq:Ei},
\end{align}
where we defined 
\begin{align}
\log^*(z)=
\begin{cases}
 \log(|z|) &\text{for } z\in\mathbb{R}\\
 \log(|z|)\mp  \frac{\i\pi}{2}&\text{for } z\in \i\mathbb{R}^\pm ,
\end{cases}
\end{align}
and \(\i\mathbb{R}^\pm\) denote the positive and negative imaginary axis respectively. We now insert this expansion into \(J(x)\), and calculate the required  sums separately. Subsequently we sort the powers of \(x\) in
\begin{align}
\begin{split}
\sum_{n=1}^\infty \frac{J(n x)}{n^2} =\sum_{n=1}^\infty \frac{1}{ n^2}
 \Bigg[\gamma\left(\text{e}^{xn}+\text{e}^{-xn}\right) +\text{e}^{xn}\log^*(-x n)+\text{e}^{-xn}\log^*(x n) 
\left.
+\text{e}^ {-x n}\sum_{k=1}^\infty \frac{\left( x n\right)^k}{k k!}+\text{e}^ {x n}\sum_{k=1}^\infty \frac{\left(- x n\right)^k}{k k!}\right]& \label{eq:sumofseries}.
\end{split}
\end{align}
In the expansion, the Polylogarithm function \(\text{Li}_s(y)\) plays an essential role. In Appendix \ref{app:polylog}
the series used for the expansion of this function are derived, with the help of the results presented there we find
\begin{align}
\sum_{n=1}^\infty \frac{\text{e}^{-xn}}{n^2}=&\text{Li}_2 \left(\text{e}^{-x}\right)=\frac{\pi^2}{6}+\mathcal{O}(x),\\
\begin{split}
\sum_{n=1}^\infty \frac{\text{e}^{-xn}}{n^2}\log(n)=&-\left.\partial_y\text{Li}_y\left(\text{e}^{-x}\right)\right|_{y=2}
=-\zeta'(2)+\mathcal{O}(x)
\end{split},\\
\begin{split}
\sum_{n=1}^\infty \frac{\text{e}^{-xn}}{n^2}\sum_{k=1}^\infty \frac{\left( x n\right)^k}{k k!}=&\sum_{k=1}^\infty \text{Li}_{2-k} \left(\text{e}^x\right)\frac{ x^k}{k k!}=\mathcal{O}(x)
\end{split},
\end{align}
We thus find to logarithmic order in \(x\) 
\begin{align}
\sum_{n=1}^\infty \frac{J(n x)}{n^2}=-2\zeta'(2)+\gamma \frac{\pi^2}{3}+\frac{\pi^2}{6}\left[\log^*(x)+\log^*(-x)\right]+\mathcal{O}(x). \label{eq:nsquaresumfull}
\end{align}
We are now interested in the sum
\begin{align}
\sum_{n=1}^\infty \frac{J(n x)}{n}\coth(z n). \label{eq:cothsum}
\end{align}
to calculate it we write  \(\coth(\pi z n)\) as
\begin{align}
 \coth(\pi z n)=1+\frac{2}{\text{e}^{2\pi nz}-1}=\text{sgn}(z)\left(1+2 \sum_{j=1}^\infty \text{e}^{-2 \pi j n \left|z\right|}\right) \label{eq:coth},
\end{align}
where the absolute value and the sign have to be introduced to ensure convergence of the sum for negative values of \(z\).
To find the result for (\ref{eq:cothsum}) we calculate the following sums using again the expansions of Appendix \ref{app:polylog} for the Polylogarithm
\begin{align}
\begin{split}
\sum_{n=1}^\infty\frac{\text{e}^{-xn}}{n}=&\text{Li}_1\left(\text{e}^{-x}\right)
=- \log(x)+\mathcal{O}(x),
\end{split}\\
\begin{split}
\sum_{n=1}^\infty \frac{\text{e}^{-xn}}{n}\log(n)=&-\left.\partial_s\text{Li}_s\left(\text{e}^{-x}\right)\right|_{s=1},
\\=&\frac{\pi^2}{12}+\frac{\gamma^2}{2}+\gamma_1+\frac{1}{2}\log(x)^2+\gamma\log(x)+\mathcal{O}(x),
\end{split}  \\
\begin{split}
\sum_{n=1}^\infty \frac{\text{e}^{-xn}}{n}\sum_{k=1}^\infty \frac{\left( x n\right)^k}{k k!}=&\sum_{k=1}^\infty \text{Li}_{1-k} \left(\text{e}^x\right)\frac{ x^k}{k k!}
\sum_{k=1}^\infty  \left(\frac{1}{k^2}+ \mathcal{O}(x)\right)=\frac{\pi^2}{6}+\mathcal{O}(x)
\end{split}.
\end{align}
So that we arrive at
\begin{align}
\begin{split}
 \sum_{n=1}^{\infty} \frac{J(n x)}{n}=&\gamma^2+2\gamma_1+\log(x)\log(-x)-\log^*(x)\log(x)-\log^*(-x)\log(-x)+\mathcal{O}(x) .\label{eq:JsumP1}
\end{split}
\end{align}
 For the second part of \(\coth(x)\) in (\ref{eq:coth})  we need
\begin{align}
\begin{split}
\sum_{n=1}^\infty\frac{\text{e}^{-  xn}}{n}\sum_{j=1}^\infty \text{e}^{-2\pi j n z}=&-\sum_{j=1}^\infty \log(1-\text{e}^{-x-2 \pi j   z})\\
=&- \log\left( \prod_{j=1}^\infty \left[ 1-\text{e}^{-2 \pi j  z}\right]\right)+\mathcal{O}(x)
\\=& -\frac{\pi}{12}z-\log\left(\eta(\i z)\right)+\mathcal{O}(x) ,
\end{split}
\end{align}
where we used the fact that \(\text{Li}_1(z)=-\log(1-z)\) (see \cite{Prudnikov2003}, Eq.~II.5) and introduced the Dedekind Eta function
\begin{align}
\eta(x)=\e^{\frac{\pi}{12} \i x}\prod_{j=1}^\infty \left(1-\e^{2 \pi \i j x }\right). \label{eq:eta}
\end{align}
This function and the identity (\ref{eq:etaidentity}) proved in \cite{Siegel1954} play a crucial role in the formulation of a fractional QED for the case studied in Section \ref{sec:Pbig}. We also need
\begin{align}
\begin{split}
\sum_{n=1}^\infty \frac{\text{e}^{-xn}}{n}\log(n)\sum_{j=1}^\infty \text{e}^{-2\pi j n  z}=&-\sum_{j=1}^\infty\left.\partial_y\text{Li}_y\left(\text{e}^{-x-2\pi j  z}\right)\right|_{y=1}
\\=&-\sum_{j=1}^\infty\left.\partial_y\text{Li}_y\left(\text{e}^{-2\pi j  z}\right)\right|_{y=1}  +\mathcal{O}(x).
\end{split}
\end{align}
Unfortunately we have not been able to perform this sum over \(j\) but, since we are primarily interested in the logarithmic growth we will ignore that sum for the time being. Furthermore we find
\begin{align}
\begin{split}
\sum_{n=1}^\infty \frac{\text{e}^{-xn}}{n}\sum_{k=1}^\infty \frac{\left( x n\right)^k}{k k!}\sum_{j=1}^\infty \text{e}^{-2\pi j n z}=&\sum_{k=1}^\infty \sum_{j=1}^\infty \text{Li}_{1-k} \left(\text{e}^{x-2\pi j  z}\right)\frac{ x^k}{k k!}\\
=&\sum_{k=1}^\infty  \sum_{j=1}^\infty \left( \text{Li}_{1-k} \left(\text{e}^{-2\pi j  z}\right)\frac{ x^k}{k k!}+ \mathcal{O}(x^{k+1})\right)\\
=&\mathcal{O}(x).
\end{split}
\end{align}
So that we find
\begin{align}
\begin{split}
  \sum_{n=1}^{\infty} \frac{J(n x)}{n}\sum_{j=1}^\infty \text{e}^{-2\pi j n  z} =&-\frac{\pi }{6}\gamma z-2 \gamma \log\left(\eta(\i z)\right)-2\sum_{j=1}^\infty\left.\partial_y\text{Li}_y\left(\text{e}^{-2\pi j  z}\right)\right|_{y=1}\\&   -\left[\frac{\pi}{12}z +\log\left(\eta(\i z)\right)\right] \left[\log^*(-x)+\log^*(x)\right]+\mathcal{O}(x)\label{eq:JsumP2}.
\end{split}
\end{align}
We can now use the fact that \(\epsilon\) and \(\beta\) are either real or purely imaginary to simplify the above results. We are just interested in \(J(x)\) and \(J(\i x)\) for real \(x\), in this case the following holds
\begin{align}
\log^*(x)+\log^*(-x)=\log^*(\i x)+\log^*(-\i  x)&=2\log(|x|), \label{eq:aslog}\\
\log(x)\log(-x)-\log^*(x)\log(x)-\log^*(-x)\log(-x)&=-\log^2(|x|),\\
\log(\i x)\log(- \i x)-\log^*(\i x)\log(\i x)-\log^*(-\i x)\log(-\i x)&=-\log^2(|x|)+\frac{3}{4}\pi^2.
\end{align} 
Now we use (\ref{eq:nsquaresumfull}) as well as (\ref{eq:JsumP1}) and (\ref{eq:JsumP2}) and put them into (\ref{eq:coth}) to find the logarithmic growth of the sums coming from the main part for real or purely imaginary \(x\) and real \(z\)
\begin{align}
\left\{\sum_{n=1}^\infty \frac{J(n x)}{n^2}\right\}_{\log}=&\frac{\pi^2}{3} \log(|x|) \label{eq:nsquaresum},\\
\begin{split}
\left\{\sum_{n=1}^\infty \frac{J(n x)}{n}\coth(\pi z n)\right\}_{\log}=& 
\left(-\frac{\pi}{3}z  -\text{sgn}(z) 4  \log\left(\eta(\i \left|z\right|)\right) \right) \log(|x|)\\&-\text{sgn}(z)\log(|x|)^2. \label{eq:cothsumlog}
\end{split}
\end{align}
\label{app:scalarsums}
For the scalar case we need the following two sums
\begin{align}
 \sum_{n=1}^\infty \frac{J(n x)}{n^2} (-1)^n \hspace{1cm}\text{and}\hspace{1cm}\sum_{n=1}^\infty \frac{J(n x)}{n^2} (-1)^n \csch(\pi z n).
\end{align}
These can be found, up to order \(x\), by using the series representation of \(\text{Ei}(x)\) analogous to (\ref{eq:sumofseries}) together with the following sums
\begin{align}
\sum_{n=1}^\infty \frac{\text{e}^{-xn}}{n^2}(-1)^n=&\text{Li}_2 \left(-\text{e}^{-x}\right)=-\frac{\pi^2}{12}+\mathcal{O}(x),\\
\begin{split}
\sum_{n=1}^\infty \frac{\text{e}^{-xn}}{n^2}\log(n) (-1)^n=&-\left.\partial_y\text{Li}_y\left(-\text{e}^{-x}\right)\right|_{y=2}
=\frac{1}{2}\zeta'(2)+\frac{\pi^2}{12}\log(2)+\mathcal{O}(x)
\end{split},\\
\begin{split}
\sum_{n=1}^\infty \frac{\text{e}^{-xn}}{n^2}(-1)^n\sum_{k=1}^\infty \frac{\left( x n\right)^k}{k k!}=&\sum_{k=1}^\infty \text{Li}_{2-k} \left(\text{e}^x\right)\frac{ x^k}{k k!}=\mathcal{O}(x)
\end{split},
\end{align}
where the series used for the Polylogarithm are derived in Appendix \ref{app:scalarpolylog}. So that we find 
\begin{align} 
\sum_{n=1}^\infty \frac{J(n x)}{n^2} (-1)^n=\zeta'(2)+\frac{\pi^2}{6}(\log(2)-\gamma)-\frac{\pi^2}{12}\left[\log^*(x)+\log^*(-x)\right]+\mathcal{O}(x). \label{eq:nsquaresumscalar}
\end{align}
By analogy with (\ref{eq:coth}), we write \(\csch(\pi z n)\) as
\begin{align}
 \csch(\pi z n)=\frac{2 \text{e}^{z}}{\text{e}^{2\pi nz}-1}=2\,\text{sgn}(z) \sum_{j=0}^\infty \text{e}^{-2 \pi \left(j+\frac{1}{2}\right) n \left|z\right|}.  \label{eq:csch}
\end{align}
We now need the sums
\begin{align}
\begin{split}
\sum_{n=1}^\infty\frac{\text{e}^{-  xn}}{n}(-1)^n\sum_{j=0}^\infty \text{e}^{-2\pi \left(j+\frac{1}{2}\right) n  z}=&-\sum_{j=0}^\infty \log\left(1+\text{e}^{-x-2 \pi \left(j+\frac{1}{2}\right)  z}\right)\\
=&- \log\left( \prod_{j=0}^\infty \left[ 1+\text{e}^{-2 \pi \left(j+\frac{1}{2}\right)  z}\right]\right)+\mathcal{O}(x) \label{eq:reasonforkappa}
\\=& \frac{\pi}{24}z-\log\left(\kappa(\i z)\right)+\mathcal{O}(x) , 
\end{split}
\end{align}
where we again used \(\text{Li}_1(z)=-\log(1-z)\) (see \cite{Prudnikov2003}, Eq.~II.5). We also introduced the function
\begin{align}
\kappa(x):=\e^{-\frac{\pi}{24} \i x}\prod_{j=1}^\infty \left(1+\e^{2 \pi \i \left(j+\frac12\right) x }\right). \label{eq:kappa}
\end{align}
Observe that unlike the Dedekind Eta function (\ref{eq:eta}) used for the spinor case the function \(\kappa(x)\) is not defined in the literature, but constructed for our purposes. However it is possible to derive the property (\ref{eq:kappaidentity}) which is analogous to the identity (\ref{eq:etaidentity}) for the Dedekind Eta function  used in section \ref{sec:Pbig}. Additionally we find
\begin{align}
\begin{split}
\sum_{n=1}^\infty \frac{\text{e}^{-xn}}{n}\log(n)(-1)^n\sum_{j=0}^\infty \text{e}^{-2\pi \left(j+\frac{1}{2}\right) n  z}=&-\sum_{j=0}^\infty\left.\partial_y\text{Li}_y\left(-\text{e}^{-x-2\pi \left(j+\frac{1}{2}\right)  z}\right)\right|_{y=1}
\\=&-\sum_{j=0}^\infty\left.\partial_y\text{Li}_y\left(-\text{e}^{-2\pi \left(j+\frac{1}{2}\right)  z}\right)\right|_{y=1}  +\mathcal{O}(x),
\end{split}
\end{align}
where we again have not been able to perform the sum over \(j\) but can see that there is no logarithmic contribution coming from this sum.
The last sum we need is
\begin{align}
\begin{split}
\sum_{n=1}^\infty \frac{\text{e}^{-xn}}{n}(-1)^n\sum_{k=1}^\infty \frac{\left( x n\right)^k}{k k!}\sum_{j=0}^\infty \text{e}^{-2\pi \left(j+\frac{1}{2}\right) n z}=&\sum_{k=1}^\infty \sum_{j=0}^\infty \text{Li}_{1-k} \left(-\text{e}^{x-2\pi \left(j+\frac{1}{2}\right)  z}\right)\frac{ x^k}{k k!}\\
=&\sum_{k=1}^\infty  \sum_{j=0}^\infty \left( \text{Li}_{1-k} \left(-\text{e}^{-2\pi \left(j+\frac{1}{2}\right)  z}\right)\frac{ x^k}{k k!}+ \mathcal{O}(x^{k+1})\right)\\
=&\mathcal{O}(x).
\end{split}
\end{align}
Combining these three sums we find
\begin{align}
\begin{split}
 \sum_{n=1}^\infty \frac{J(n x)}{n^2} (-1)^n \csch(\pi z n)=&\frac{\pi}{6}\gamma z-4\,\text{sgn}(z)\gamma \left[\log\left(\kappa(\i |z|)\right)+\sum_{j=0}^\infty\left.\partial_y\text{Li}_y\left(-\text{e}^{-2\pi \left(j+\frac{1}{2}\right)  z}\right)\right|_{y=1}\right]\\&+ \left(\frac{\pi}{12}z-\text{sgn}(z)2\log\left(\kappa(\i |z|)\right)\right) \left[\log^*(x)+\log^*(-x)\right]+\mathcal{O}(x).
\end{split}
\end{align}
So that in the end using the restrictions on \(\epsilon\) and \(\beta\) and (\ref{eq:aslog}) we find the logarithmic growth of the sums for real or purely imaginary \(x\) and real \(z\)
\begin{align}
\left\{\sum_{n=1}^\infty (-1)^n \frac{J(n x)}{n^2}\right\}_{\log}=&-\frac{\pi^2}{6} \log(|x|) ,\\
\begin{split}
\left\{\sum_{n=1}^\infty (-1)^n \frac{J(n x)}{n}\csch(\pi z n)\right\}_{\log}=&\left(\frac{\pi}{6}z-4\,\text{sgn}(z)\log\left(\kappa(\i |z|)\right)\right)  \log(|x|) \label{eq:cschsumlog}.
\end{split}
\end{align}

\section{Polylogarithms}
\label{app:polylog}
Polylogarithms are defined by (see e.g. \cite{Wood1992})
\begin{align}
 \text{Li}_s(z):=\sum_{j=1}^\infty \frac{x^j}{j^s}=\frac{1}{\Gamma(s)}\int_0^\infty \frac{t^{s-1}}{\frac{\e^t}{z}-1}dt.
\end{align}
for all complex \(z,\,s\) except \(z\) real and \(z>1\), where \(\Gamma(s)\) is the Gamma function. The integral representation only holds for \(\text{Re}(s)>0\) but can be extended to negative values with the method of contour integration \cite{Whittaker1996} 
\begin{align}
\text{Li}_s\left(\e^x\right)=\frac{\Gamma(1-s)}{2\pi \i}\int_H \frac{(-t)^{s-1}}{\e^{t-\mu}-1}dt, \label{eq:hankel}
\end{align}
where \(H\) is the Hankel contour which starts at \(t=\infty\) in the upper half of the  complex plane goes to \(t=0\) encircles the origin in a counterclockwise sense and then runs to  \(t=\infty\) in the lower half of the complex plane. It is now possible to modify the contour to enclose the poles and to evaluate the integral as the sum of residues
\begin{align}
\text{Li}_s\left(\e^x\right)=\Gamma(1-s)\sum_{k=-\infty}^\infty(2 k \i \pi  -\mu)^{p-1}. \label{eq:Lisum}
\end{align}
By expanding in \(\mu\) and summing over \(s\) we find the series representation (see \cite{Wood1992}, Eq.~(9.3))
\begin{align}
 \text{Li}_s(\text{e}^x)=\Gamma(1-s)(-x)^{s-1}+\sum_{j=0}^\infty \frac{\zeta(s-j)}{j!}x^j, \label{eq:polysum1}
\end{align}
where \( \zeta(z)\) is the Riemann-Zeta function. \\ 
For positive integer values \(s=n=1,2,3\) both \(\Gamma(1-n)\) and \(\zeta(1)\) in the sum (for \(j=n+1\)) diverge. If we however look at the expansion around the poles for the respective functions (see \cite{Prudnikov2003}, Eq. II.3. and \cite{Havil2003}, p.~118)
\begin{align}
 \Gamma(1-n-\delta)&=\frac{(-1)^n}{(n-1)!}\left[\frac{1}{\delta}-\psi(n)+\frac{\delta}{2}\left(\psi(n)^2-\psi'(n)+\frac{\pi^2}{3}\right)\right]+\mathcal{O}(\delta^2) \label{eq:gammasum},\\
 \zeta(1+\delta)&=\frac{1}{\delta}+\gamma-\delta\gamma_1+\mathcal{O}(\delta^2) \label{eq:zetasum},
\end{align}
where \(\gamma_1\) is the first Stieltjes constant defined by
\begin{align}
 \gamma_1=\lim_{n\rightarrow\infty}\left(\sum_{k=1}^n \frac{\log(k) }{k}-\frac{\log(n)^2}{2}\right)\approx0.577,
\end{align}
and the expansion
\begin{align}
 (-x)^{n+\delta-1}=(-x)^{n-1}\left(1+ \log(-x)\delta+\frac{1}{2}\log(-x)^2\delta^2+\mathcal{O}(\delta^3)\right),
\end{align}
we find
\begin{align}
\lim_{\delta \rightarrow 0 }\left[\Gamma(1-n-\delta)+\frac{\zeta(1+\delta) x^{n-1}}{(n-1)!}\right]=\frac{ x^{n-1}}{(n-1)!}\left[\psi(n)+\gamma-\log(-x)\right].
\end{align}
So that for positive integers we find the summation formula (see also \cite{Wood1992}, Eq.~(9.5))
\begin{align}
 \text{Li}_n(\text{e}^x)=\frac{ x^{n-1}}{(n-1)!}\left[\psi(n)+\gamma-\log(-x)\right]+\sum_{j=0,j\ne n-1}^\infty \frac{\zeta(n-j)}{j!}x^j. \label{eq:polysum2}
\end{align}
We also need series for the derivatives of \(\text{Li}_s(\e^x)\) with respect to \(s\) for \(s=n=1,2\). We could not find those in the literature but it is possible to derive them using the definition of the derivative and the formulas above
\begin{align}
 \left. \partial_s \text{Li}_s(\text{e}^x) \right|_{s=n}= \lim_{\delta \rightarrow 0} \frac{\text{Li}_{n+\delta}(\text{e}^x)- \text{Li}_{n}(\text{e}^x)}{\delta},
\end{align}
where for the first Polylogarithm we have to use (\ref{eq:polysum1}) and for the second (\ref{eq:polysum2})
\begin{align}
\begin{split}
 \left. \partial_s \text{Li}_s(\text{e}^x) \right|_{s=n}=& \lim_{\delta \rightarrow 0} \frac{1}{\delta}\left\{\Gamma(1-n-\delta)(-x)^{s+\delta-1}+\frac{ x^{n-1}}{(n-1)!}\zeta(1+\delta) 
- \frac{ x^{n-1}}{(n-1)!}\left[\psi(n)+\gamma-\log(-x)\right]\right\}\\&+\sum_{j=0,j\ne n-1}^\infty \frac{\zeta'(s-j)}{j!}x^j,
\end{split}
\end{align}
we can now again use the expansion of Gamma and Zeta function around the poles (\ref{eq:gammasum}) and (\ref{eq:zetasum}) respectively to find
\begin{align}
\begin{split}
  \left. \partial_s \text{Li}_s(\text{e}^x) \right|_{s=n}=& \frac{ x^{n-1}}{(n-1)!}\left[-\frac12 \log(-x)^2+\psi(n)\log(-x)
-\frac{1}{2}\left(\psi(n)^2-\psi'(n)\right)-\frac{\pi^2}{6}-\gamma_1\right]
\\&
+\sum_{j=0,j\ne n-1}^\infty \frac{\zeta'(n-j)}{j!}x^j. \label{eq:polyder}
\end{split}
\end{align}
The Polylogarithms we need in Appendix \ref{app:sum2} have the following expansions up to order 1 in \(x\) for \(k>1\)
\begin{align}
\text{Li}_1(\e^{-x})=&-\log(x)+\mathcal{O}(x),\\
\text{Li}_2(\e^{-x})=&\frac{\pi^2}{6}+\mathcal{O}(x),\\
(x)^k \text{Li}_{1-k}(\e^{-x})=&\Gamma[k]+\mathcal{O}(x),\\
(x)^k \text{Li}_{2-k}(\e^{-x})=&\mathcal{O}(x),\\
\begin{split}
 \partial_s\left. \text{Li}_s(\e^{-x})\right|_{s=1}=&-\frac{\pi^2}{12}-\frac{\gamma^2}{2}-\gamma_1-\log(x)^2-\gamma\log(x)+\mathcal{O}(x)
\end{split}\\
\partial_s\left.\text{Li}_s(\e^{-x})\right|_{s=2}=&\zeta'(2)+\mathcal{O}(x),
\end{align}
where we have used the known values \(\psi(1)=-\gamma\) and \(\zeta(2)=\psi'{(1)}=\pi^2/6\) (see \cite{Grad2000}, Eqs.~8.366 and 9.542).

\label{app:scalarpolylog}
For the calculations for scalar QED we are interested in \(\text{Li}_s\left(-\e^x\right)\). By analogy with (\ref{eq:Lisum}) we find
\begin{align}
\text{Li}_s\left(-\e^x\right)=\Gamma(1-s)\sum_{k=-\infty}^\infty((2 k-1) \i \pi -\mu)^{p-1}.
\end{align}
which by again expanding in \(\mu\) and summing over \(k\) can be brought in the form 
\begin{align}
 \text{Li}_s(-\text{e}^x)=-\sum_{j=0}^\infty \frac{\zeta(s-j)}{j!}x^j(1-2^{j+1-s}), 
\end{align}
for positive integer values \(s=n=1,2,3,\dots\) and \(j=n-1\). The sum contains the divergent \(\zeta(1)\), however it is finite since
\begin{align}
 \lim_{j\rightarrow n-1}(1-2^{j+1-s})\zeta(s-j)=-\log(2).
\end{align}
Such that we find
\begin{align}
 \text{Li}_n(-\text{e}^x)=-\log(2)\frac{x^{n-1}}{(n-1)!}-\sum_{j=0,j\ne n-1}^\infty \frac{\zeta(n-j)}{j!}x^j (1-2^{j+1-n}). 
\end{align}
We also need the derivative of the Polylogarithm with respect to \(s\) for positive integer values which can be derived analogously to (\ref{eq:polyder}) and takes the form
\begin{align}
\begin{split}
  \left. \partial_s \text{Li}_s(-\text{e}^x) \right|_{s=n}=&\left(\frac{1}{2}\log(2)^2-\gamma\log(2)\right) \frac{ x^{n-1}}{(n-1)!}\\-&\sum_{j=0,j\ne n-1}^\infty \frac{x^j}{j!}\left(\zeta'(n-j)(1-2^{j+1-n})+\log(2)2^{j+1-n}\zeta(n-j)\right).
\end{split}
\end{align}
So that the Polylogarithms needed in Appendix \ref{app:scalarsums} take the form
\begin{align}
\text{Li}_2(-\e^{-x})=&-\frac{\pi^2}{12}+\mathcal{O}(x),\\
(x)^k \text{Li}_{2-k}(-\e^{-x})=&\mathcal{O}(x),\\
\partial_s\left.\text{Li}_s(-\e^{-x})\right|_{s=2}=&-\frac{1}{2}\zeta'(2)-\frac{\pi^2}{12}\log(2)+\mathcal{O}(x),
\end{align}
where we have used \(\zeta(2)=\pi^2/6\) (see \cite{Grad2000}, Eq.~9.542).

\section{Identity for the function \texorpdfstring{\(\kappa(x)\)}{kappa(x)}}
\label{app:kappa}
In this appendix we prove an identity for the  function \(\kappa(x)\) which is useful for our calculations. The proof is inspired by a proof for an identity for the Dedekind Eta function \cite{Siegel1954}.
We will use 
\begin{align}
 f(z)=\csc(z)\csc\left(\frac{z}{\tau}\right),\hspace{1cm} \nu=\left(n+\frac12\right)\pi \hspace{1cm} n\in \mathbb{N}, \hspace{1cm} \i\tau\in\mathbb{R} .
\end{align}
The function \(F(z)=z^{-1}f(\nu z)\) has poles of order one at 
\begin{align*}
&  z=\pm\frac{\pi k}{\nu} & &\text{ with residue }& \frac{(-1)^k}{\pi k}\cot \left(\frac{\pi k}{\tau}\right)\\
\intertext{and}
&  z=\pm\frac{\pi k \tau}{\nu} & &\text{ with residue }& \frac{(-1)^k}{\pi k}\cot \left({\pi k}{\tau}\right)
\intertext{for \(k\in \mathbb{N}\) as well as one pole of order three at}
&z=0 & &\text{ with residue }& \frac{1}{6}\left(\tau+\frac{1}{\tau}\right).
\end{align*}
If we now choose the contour \(C_\lozenge\) to be the rhombus with endpoints at \(1,\tau,-1,\tau\) in the \(z\)-plane we can use the residue theorem to find
\begin{align}
 \int_{C_\lozenge}f(\nu z)g(\nu a z)\frac{dz}{z}=4i\sum_{k=1}^n \frac{(-1)^k}{k}\left[\csc \left(\frac{\pi k}{\tau}\right)+\csc \left({\pi k}{\tau}\right)\right]+\frac{\pi \i}{3}\left(\tau+\frac{1}{\tau}\right) \label{eq:nobranchcut}.
\end{align}
Now we can write the co-secant function as
\begin{align}
 \csc(x)=2\i\frac{\text{e}^{-2\i x}}{1-\text{e}^{-2\i x}}=-2\i\frac{\text{e}^{2\i x}}{1-\text{e}^{2\i x}},
\end{align}
which shows that the limit of \(f(\nu z)\) for \(n\rightarrow \infty\) everywhere in the complex plane except the axes is 0. We thus find
\begin{align}
\sum_{k=1}^n \frac{(-1)^k}{k}\left[\csc \left(\frac{\pi k}{\tau}\right)+\csc \left({\pi k}{\tau}\right)\right]=-\frac{\pi}{12}\left(\tau+\frac{1}{\tau}\right) \label{eq:cotsums}
\end{align}
Observe that for \(\tau=\i\epsilon/\beta\) this  is proportional to the constant order of (\ref{eq:scalarEHRe}). We are however mainly interested in the logarithmic order. One could now try to set \(F(z)=z^{-1}f(\nu z)\ln(\nu z E_c/\epsilon)\), which would in fact give the logarithmic order of the sum in the residues. However things get more complicated due to the branch cut of the logarithm. More importantly the integral on the contour \(C_\lozenge\) diverges.

Finally it is possible to start from \(F(z)=z^{-1}f(\nu z)J(\nu z E_c/\epsilon)\). By doing so, the sum of the residues gives exactly (\ref{eq:scalarfull}) and the integral along the contour \(C_\lozenge\) vanishes again. One however has to alter the integration contour to avoid the branch cuts. Integrating along the branch cuts gives exactly the integral (\ref{eq:EHscalar}). This is not surprising since this is the integral we started from to derive (\ref{eq:scalarfull}).

We can use (\ref{eq:cotsums}) to find an identity for the function \(\kappa(x)\) which we defined in analogy to the Dedekind Eta function in Appendix \ref{app:scalarsums}. To do so we use the definition of \(\kappa(x)\) (\ref{eq:kappa}) and (\ref{eq:csch}) to find
\begin{align} 
\begin{split}
\sum_{k=1}^{\infty}\frac{(-1)^k}{k}\csc(\i\pi k z)=&-\i\sum_{k=1}^{\infty}\frac{(-1)^k}{k}2\,\text{sgn}(z) \sum_{j=0}^\infty \text{e}^{-2 \pi \left(j+\frac{1}{2}\right) n \left|z\right|k}. \\
=&-2\i\,\text{sgn}(z) \sum_{j=0}^\infty \log\left[1+ \text{e}^{-2 \pi \left(j+\frac{1}{2}\right) n \left|z\right|k}\right]\\
=&-2\i\,\text{sgn}(z) \log[\kappa(\i |z|)]-\frac{\pi}{12}\i z
 \end{split}
\end{align}
This again is not surprising since the need to define \(\kappa(x)\) arose from the summation over the hyperbolic co-secant function in equation (\ref{eq:reasonforkappa}). Now using this and (\ref{eq:cotsums}) we arrive at the identity
\begin{align}
 \kappa\left(-\frac {1}{\tau}\right)=\kappa(\tau) 
\label{eq:kappaidentity}
\end{align}
for \(\tau\) purely imaginary. Observe that although the definition of \(\kappa(x)\) may seem artificial it was defined for convenience and to show parallels to the spinor case. It would of course be possible to do the whole calculations without defining the function and use directly (\ref{eq:cotsums}) instead of the identity. 




\end{document}